\documentclass[twocolumn]{aastex63}

\usepackage{soul}
\usepackage{amsmath}

\received{April 20, 2021}
\submitjournal{ApJ Letters}
\accepted{July 25, 2021}

\shortauthors{Goncharov et al.}

\graphicspath{{./}{}}

\begin{document}


\title{On the evidence for a common-spectrum process in the search for the nanohertz gravitational-wave background with the Parkes Pulsar Timing Array}
\shorttitle{Evidence for a common-spectrum process}


\author{Boris Goncharov}%
 \email{boris.goncharov@me.com}
\affiliation{Centre for Astrophysics and Supercomputing, Swinburne University of Technology, PO Box 218, Hawthorn, VIC 3122, Australia}
\affiliation{ARC Centre of Excellence for Gravitational Wave Discovery}
\affiliation{Gran Sasso Science Institute (GSSI), I-67100 L'Aquila, Italy}

\author{R. M. Shannon}
\affiliation{Centre for Astrophysics and Supercomputing, Swinburne University of Technology, PO Box 218, Hawthorn, VIC 3122, Australia}
\affiliation{ARC Centre of Excellence for Gravitational Wave Discovery}

\author{D.~J.~Reardon}
\affiliation{Centre for Astrophysics and Supercomputing, Swinburne University of Technology, PO Box 218, Hawthorn, VIC 3122, Australia}
\affiliation{ARC Centre of Excellence for Gravitational Wave Discovery}

\author{G.~Hobbs}
\affiliation{Australia Telescope National Facility, CSIRO, Space and Astronomy, PO Box 76, Epping, NSW 1710, Australia}

\author{A.~Zic}
\affiliation{Department of Physics and Astronomy, and Research Centre in Astronomy, Astrophysics and Astrophotonics, Macquarie University, NSW 2109, Australia}
\affiliation{Australia Telescope National Facility, CSIRO, Space and Astronomy, PO Box 76, Epping, NSW 1710, Australia}


\author{M.~Bailes}
\affiliation{Centre for Astrophysics and Supercomputing, Swinburne University of Technology, PO Box 218, Hawthorn, VIC 3122, Australia}
\affiliation{ARC Centre of Excellence for Gravitational Wave Discovery}

\author{M.~Cury\l{}o}
\affiliation{Astronomical Observatory, University of Warsaw, Al. Ujazdowskie 4, 00-478 Warsaw, Poland}

\author{S.~Dai}
\affiliation{Australia Telescope National Facility, CSIRO, Space and Astronomy, PO Box 76, Epping, NSW 1710, Australia}
\affiliation{Western Sydney University, Locked Bag 1797, Penrith South DC, NSW 1797, Australia}

\author{M.~Kerr}
\affiliation{Space Science Division, Naval Research Laboratory, Washington, DC 20375-5352, USA}

\author{M.~E.~Lower}
\affiliation{Centre for Astrophysics and Supercomputing, Swinburne University of Technology, PO Box 218, Hawthorn, VIC 3122, Australia}
\affiliation{Australia Telescope National Facility, CSIRO, Space and Astronomy, PO Box 76, Epping, NSW 1710, Australia}

\author{R.~N.~Manchester}
\affiliation{Australia Telescope National Facility, CSIRO, Space and Astronomy, PO Box 76, Epping, NSW 1710, Australia}

\author{R.~Mandow}
\affiliation{Australia Telescope National Facility, CSIRO, Space and Astronomy, PO Box 76, Epping, NSW 1710, Australia}
\affiliation{Centre for Astrophysics and Supercomputing, Swinburne University of Technology, PO Box 218, Hawthorn, VIC 3122, Australia}

\author{H.~Middleton}
\affiliation{Centre for Astrophysics and Supercomputing, Swinburne University of Technology, PO Box 218, Hawthorn, VIC 3122, Australia}
\affiliation{School of Physics, University of Melbourne, Parkville, VIC 3010, Australia}
\affiliation{ARC Centre of Excellence for Gravitational Wave Discovery}

\author{M.~T.~Miles}
\affiliation{Centre for Astrophysics and Supercomputing, Swinburne University of Technology, PO Box 218, Hawthorn, VIC 3122, Australia}
\affiliation{ARC Centre of Excellence for Gravitational Wave Discovery}

\author{A.~Parthasarathy}
\affiliation{Max-Planck-Institut f\"{u}r Radioastronomie, Auf dem H\"{u}gel 69, D-53121 Bonn, Germany}

\author{E.~Thrane}
\affiliation{School of Physics and Astronomy, Monash University, Clayton, VIC 3800, Australia}
\affiliation{ARC Centre of Excellence for Gravitational Wave Discovery}

\author{N.~Thyagarajan}
\affiliation{Australia Telescope National Facility, CSIRO, Space and Astronomy, PO Box 76, Epping, NSW 1710, Australia}

\author{X.~Xue}
\affiliation{CAS Key Laboratory of Theoretical Physics, Insitute of Theoretical Physics, Chinese Academy of Sciences, Beijing 100190, China}
\affiliation{School of Physical Sciences, University of Chinese Academy of Sciences, Beijing 100049, China}

\author{X.-J.~Zhu}
\affiliation{School of Physics and Astronomy, Monash University, Clayton, VIC 3800, Australia}
\affiliation{ARC Centre of Excellence for Gravitational Wave Discovery}



\author{A.~D.~Cameron}
\affiliation{Centre for Astrophysics and Supercomputing, Swinburne University of Technology, PO Box 218, Hawthorn, VIC 3122, Australia}
\affiliation{ARC Centre of Excellence for Gravitational Wave Discovery}



\author{Y.~Feng}
\affiliation{CAS Key Laboratory of FAST, National Astronomical Observatories, Chinese Academy of  Sciences, Beijing 100101, People's Republic of China}

\author{R.~Luo}
\affiliation{Australia Telescope National Facility, CSIRO, Space and Astronomy, PO Box 76, Epping, NSW 1710, Australia}


\author{C. J. Russell}
\affiliation{CSIRO Scientific Computing, Australian Technology Park, Locked Bag 9013, Alexandria, NSW 1435, Australia}

\author{J.~Sarkissian}
\affiliation{CSIRO  Space and Astronomy, Australia Telescope National Facility, 	PO Box 276 Parkes NSW 2870 Australia}

\author{R.~Spiewak}
\affiliation{Jodrell Bank Centre for Astrophysics, Department of Physics and Astronomy, University of Manchester, Manchester M13 9PL, UK}
\affiliation{Centre for Astrophysics and Supercomputing, Swinburne University of Technology, PO Box 218, Hawthorn, VIC 3122, Australia}


\author{S.~Wang}
\affiliation{Xinjiang Astronomical Observatory, Chinese Academy of Sciences, 150 Science 1-Street, Urumqi, Xinjiang 830011, People's Republic of China}
\affiliation{Australia Telescope National Facility, CSIRO, Space and Astronomy, PO Box 76, Epping, NSW 1710, Australia}

\author{J.~B.~Wang}
\affiliation{Xinjiang Astronomical Observatory, Chinese Academy of Sciences, 150 Science 1-Street, Urumqi, Xinjiang 830011, People's Republic of China}

\author{L.~Zhang}
\affiliation{National Astronomical Observatories, Chinese Academy of Sciences, A20 Datun Road, Chaoyang District, Beijing 100101, People’s Republic of China}

\author{S.~Zhang}
\affiliation{Purple Mountain Observatory, Chinese Academy of Sciences, Nanjing 210008, People’s Republic of China}




\begin{abstract}

A nanohertz-frequency stochastic gravitational-wave background can potentially be detected through the precise timing of an array of millisecond pulsars.
This background produces low-frequency noise in the pulse arrival times that would have a characteristic spectrum common to all pulsars and a well-defined spatial correlation.
Recently the North American Nanohertz Observatory for Gravitational Waves collaboration (NANOGrav) found evidence for the common-spectrum component in their 12.5-year data set.
Here we report on a search for the background using the second data release of the Parkes Pulsar Timing Array.
If we are forced to choose between the two NANOGrav models $––$ one with a common-spectrum process and one without $––$ we find strong support for the common-spectrum process.
However, in this paper, we consider the possibility that the analysis suffers from model misspecification.
In particular, we present simulated data sets that contain noise with distinctive spectra but show strong evidence for a common-spectrum process under the standard assumptions.
The Parkes data show no significant evidence for, or against, the spatially correlated Hellings-Downs signature of the gravitational-wave background.
Assuming we did observe the process underlying the spatially uncorrelated component of the background, we infer its amplitude to be $A = 2.2^{+0.4}_{-0.3} \times 10^{-15}$ in units of gravitational-wave strain at a frequency of $1\, \text{yr}^{-1}$.
Extensions and combinations of existing and new data sets will improve the prospects of identifying spatial correlations that are necessary to claim a detection of the gravitational-wave background.

\end{abstract}


\section{Introduction} \label{sec:intro}

While detections of gravitational waves \cite[e.g.,][]{gw150914} have been made with ground-based interferometers that are sensitive to hertz-kilohertz gravitational waves, experiments that operate at lower frequencies have yet to identify a signal.
Pulsar timing array (PTA) experiments, which monitor and measure arrival times from millisecond pulsars (MSPs),  have been established to search for signals in the nanohertz band.
This is the domain of the stochastic background from supermassive black hole binaries, which is expected to be the first gravitational-wave signal detected with PTAs~\citep{rosado2015properties}.
The background manifests as a temporally and spatially correlated process in the MSP arrival times.
The strain spectrum of such a background is predicted to have the power-law form
$h(f) = A (f/{\rm 1\, yr^{-1}})^{-2/3}$, 
where $f$ is the gravitational-wave frequency and $A$ is the strain amplitude at $f=1~\text{yr}^{-1}$ \cite[][]{phinney_backgrounds}.
The amplitude, $A$, will depend on the demographics of the supermassive black hole population.
Astrophysical effects relating to supermassive binary black hole evolution may cause deviations from a power law \cite[e.g.,][]{ravi_gwb,sampson2015finalparsec,taylor2017env}.

A definitive detection  of the gravitational wave  background is the presence of specific spatial correlations in the arrival times~\citep{hellingsdowns1983}.
Other processes can produce signals with similar temporal properties, with either no \citep{shannon_timingnoise} or different spatial correlation. \citet{tiburzicorrelations} and \citet{taylor2017allcorrelations} showed the challenges in distinguishing between sources that produce different spatial correlations.  

Previous searches for the gravitational-wave background from supermassive black hole binaries have reported limits  on the strain amplitude, $A$, ranging between $1.1\times 10^{-14}$ and $1.0 \times 10^{-15}$ at $95\%$ of either confidence or credibility, where appropriate \citep{jenet2006gwb,vanhaasteren2011gwb,demorest2013gwb,shannon2013gwb,lentati2015gwb,arzoumanian2015gwb9yr,shannon2015gwb,arzoumanian2018gwb11yr}.
The limits are now known to be affected by systematic uncertainties in the ephemeris of the solar system, which impacts pulsar timing because the arrival times are necessarily referenced to an inertial frame located at solar system barycentre \cite[e.g.,][]{arzoumanian2018gwb11yr}. 

In a recent analysis of the NANOGrav 12.5-year data set, a common noise process was reported having a Bayes factor greater than $10^4$ (this corresponds to $9$ on the commonly used natural logarithmic scale), with the signal persisting even when accounting for uncertainties in the solar system ephemeris \cite[][]{arzoumanian2020gwb}.  We discuss the meaning of the term ``common process'' in detail later. Here we define the symbol ${\rm CP}$ to represent the common process as obtained by the hypotheses used in the NANOGrav analysis. 
Evidence for Hellings-Downs correlations was  insignificant. 

We have carried out a similar analysis using the Parkes Pulsar Timing Array \cite[PPTA;][]{manchesterppta} second data release \cite[][]{kerr2020pptadr2}. The observations and methodology are described in Section~\ref{sec:data}. 
In Section~\ref{sec:results} we discuss the results of the searches. In Section~\ref{sec:discussion}, we discuss limitations in the methodology.
In particular,  we demonstrate through simulation that the search methods can spuriously detect  a common red process in timing array data sets in which it is absent.

\section{The data set and methodology} \label{sec:data}

The PPTA project monitors an ensemble of MSPs with the 64-m Parkes radio telescope (also named {\em Murriyang}) in New South Wales, Australia. The data used, namely pulse arrival times from the observations, were acquired between 2003 and 2018 and were published as part of the second data release of the project \cite[PPTA-DR2;][]{kerr2020pptadr2}.   Observations were taken at a cadence of approximately three weeks.  At each epoch, data were usually recorded in bands centred at three different radio frequencies in order to correct variations in pulsar dispersion measures \cite[][]{keithdmvariations}. Data were recorded with a series of digital processing systems, with  quality having improved over the course of the project. 


We analyzed the data set using methodology that was based on that applied to the  NANOGrav 12.5-year data set~\citep{arzoumanian2020gwb}, which itself was based on \cite{arzoumanian2015gwb9yr} and \cite{taylor2017allcorrelations}.
Stochastic signals were modeled as being correlated (red) or uncorrelated (white) in time. 
We had previously characterized the noise processes for individual pulsars in the PPTA sample \cite[][]{goncharov2020pptadr2noise}.  
That analysis showed that the PPTA data sets contain a wide variety of noise processes, including instrument-dependent or band-dependent processes.
In this work we included red-noise processes in all pulsars, even for those pulsars that showed no evidence for such noise in previous analyses.  

As in \cite{goncharov2020pptadr2noise}, we assume that the power spectral density of all red processes follows a power law, parameterized such that the amplitude, $A$, is in units of gravitational-wave strain at $1\,\text{yr}^{-1}$:
\begin{equation}\label{eq:powerlaw}
    P(f|A,\gamma) = \Gamma(\zeta_{ab}) \frac{A^2}{12 \pi^2} \bigg(\frac{f}{\text{yr}^{-1}}\bigg)^{-\gamma} \text{yr}^3.
\end{equation}
The fluctuation frequency of the pulse arrival time power spectrum is denoted $f$ and the spectral index is $-\gamma$.
The noise terms are modeled using a Fourier series, starting with a fundamental  frequency that is the inverse of the observation span corresponding to the entire pulsar data set, $T_\text{obs}$.  We use $n_\text{c} = 30$ harmonics if $\gamma > 1.5$ \cite[][]{goncharov2020turnover}, otherwise we use  $n_\text{c}$ from the single-pulsar analysis~\citep{goncharov2020pptadr2noise}.
The overlap reduction function    \cite[$\Gamma(\zeta_{ab})$;][]{finnorf} 
characterizes the spatial correlation of the signal, and depends on the angular separation, $\zeta_{ab}$, of two pulsars $a$ and $b$ with respect to the observer.
For an isotropic stochastic background from the gravitational waves of General Relativity ~\citep{hellingsdowns1983,jenet2015orf},
\begin{equation}\label{eq:hd}
\begin{split}
    \Gamma_{\text{GWB}}(\zeta_{ab}) = & \frac{1}{2} - \frac{1}{4} \bigg( \frac{1 - \cos{\zeta_{ab}}}{2} \bigg) + \\
    & \frac{3}{2} \bigg( \frac{1 - \cos{\zeta_{ab}}}{2} \bigg) \ln{\bigg(\frac{1 - \cos{\zeta_{ab}}}{2}\bigg)},
\end{split}
\end{equation}
when $a \neq b$.

Our analysis proceeded through these steps:
\begin{itemize}
    \item We first searched for a common power-law, red-spectrum stochastic process (${\rm CP1}$), with an identical power spectrum and unrelated temporal evolution or spatial correlation across pulsars. We emphasize that the timing spectrum of the process is assumed to be statistically identical ensemble-average power spectrum among pulsars, which would be the case for a gravitational wave background.\footnote{This term was first introduced in \cite{arzoumanian2018gwb11yr}, and we refer the reader to that paper for further discussion of its meaning.}    Throughout our analysis we marginalised over deterministic terms in the timing model \citep{2021arXiv210704609R}. This included instrument-dependent offsets (``jumps'') of unknown value, as identified by \cite{kerr2020pptadr2}.  Offsets with {\em a-priori} measured values were held fixed.  
    We trialed both marginalising over the white-noise parameters and also holding them fixed at their maximum {\em a-posteriori} values.   We obtained consistent results between these approaches. 
    \item Following  the NANOGrav analysis we also assumed a power-law model with $\gamma = 13/3$. This value has astrophysical interest as it is the expected value for a gravitational wave background caused by supermassive binary black holes~\citep{phinney_backgrounds}.  The resulting common power-law, red-noise process is here labeled as ${\rm CP2}$.  Based on a factorized-likelihood approach, we performed a dropout analysis to evaluate the consistency of individual PPTA DR2 pulsars with the signal identified by ${\rm CP2}$  (see \citealt{arzoumanian2020gwb}).
    \item We measured the amplitude of individual Fourier components $P_i(f_i|\rho_i) = \rho_i^2~T_\text{obs}$ of a common process, at each frequency $f_i$ separately in order to determine whether the power-law assumption of ${\rm CP1,2}$ is valid.
    \item We searched for evidence that ${\rm CP2}$ exhibits spatial correlations from a gravitational wave background, a monopolar signal, MP, or a dipolar signal, DP.  In this analysis we held the white-noise stochastic components fixed at their maximum {\em a-posteriori} value to reduce the number of parameters in the search and reduce computation time.  
    \item To assess the shape of any spatial correlations, we measured  $\Gamma(\zeta_{ab})$ at seven equally separated ``node'' angles between $0$ and $180\deg$ inclusive, using linear interpolation to determine $\Gamma(\zeta_{ab})$ from pulsar pairs between the nodes. The interpolant modeling of the PTA correlation curve was first done in \cite{taylor2013orf}. 
\end{itemize}





\subsection{Comparison with the NANOGrav data set and processing methods}

The NANOGrav analysis included the timing data from 45 pulsars in their analysis of their 12.5-year data set~\citep{arzoumanian2020gwb}. The PPTA-DR2 analysis is based on the data from 26 pulsars spanning up to 15 years.

While the data sets were obtained with different telescopes at a different range of frequencies,  the two data sets have $11$ pulsars in common, including some of the most precisely timed pulsars. 
The sources in common are PSRs~J0613$-$0200, J1024$-$0719, J1600$-$3053, J1643$-$1224, J1713$+$0747, J1832$-$0836, J1857+0943, J1909$-$3744, J1939+2134 and J2145$-$0750.   
Even for these pulsars we independently determined noise models. Our instrumental noise terms are necessarily independent from NANOGrav. 
However, we also included extra noise terms into the modelling for specific pulsars. 
In particular for PSR~J1713+0747 the NANOGrav analysis included timing noise and dispersion measure noise terms as well as the inclusion of two exponential dips attributed to rapid dispersion measure variations.  
The PPTA analysis is similar. In  the second exponential dip we allowed for a different chromaticity  as there is evidence that it is not caused  by dispersion-measure variations~\citep{goncharov2020pptadr2noise}. 

Both analyses made use of \textsc{tempo2}. 
Bayesian inference was performed with \textsc{enterprise} \cite[][]{ellis2019enterprise}. 
Preferred models were selected based on the Bayes factor calculated using a product-space sampling method \cite[][]{carlin1995bayesian,taylor2020productspace}.
We denote a Bayes factor for model $\text{A}$ over model $\text{B}$ as $\mathcal{B}^{\text{A}}_{\text{B}}$.
The null model, with no common or correlated noise processes in the data set, is denoted $\varnothing$.
We referenced pulse arrival times to the solar system barycenter using the ephemeris DE436, to maintain consistency with PPTA-DR2 \citep{kerr2020pptadr2} and the single-pulsar analyses~\citep{goncharov2020pptadr2noise,2021arXiv210704609R}.  
The more recent DE438 ephemeris was used in the NANOGrav analysis.


\section{Results} 
\label{sec:results} 

\begin{figure*}
    \includegraphics[width=0.43\textwidth]{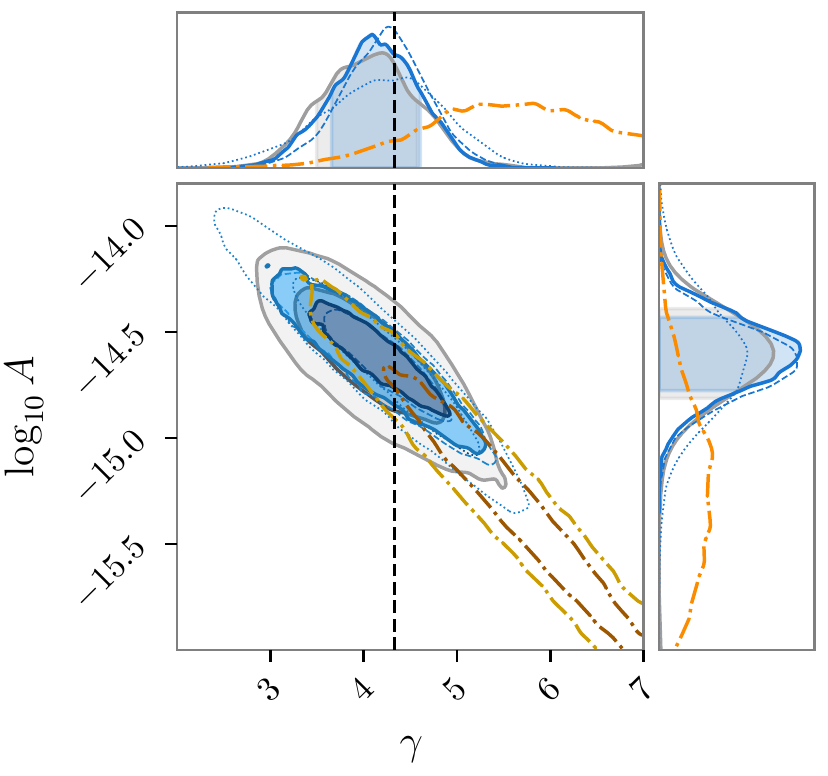}
    \includegraphics[width=0.55\textwidth]{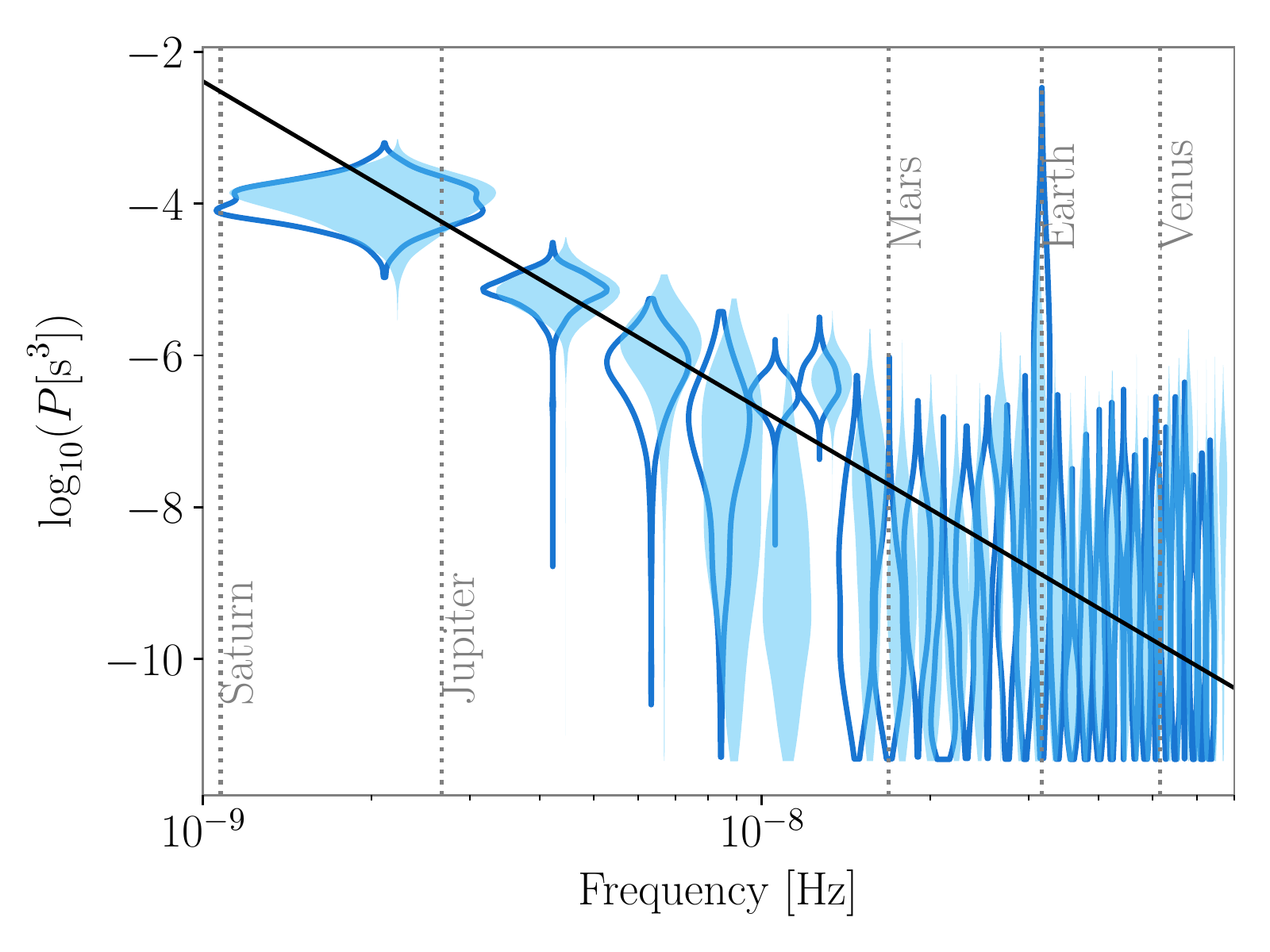} 
    \caption{Left: Measurements of common power-law red-noise parameters and the demonstration of their robustness to assumptions about pulsar-intrinsic noise and the number of fluctuation frequencies $n_\text{c}$. The dashed vertical line indicates $\gamma=13/3$.  The solid lines represent the measurement based on $n_\text{c} = 30$. Dashed and dotted lines represent $n_\text{c} = 20$ and $n_\text{c} = 5$. The dash-dotted lines correspond to the measurement from~\cite{arzoumanian2020gwb}. Contours and shaded regions are 1-$\sigma$ and 2-$\sigma$ credible levels.
    Grey lines and regions are based on the assumption of achromatic timing noise in every pulsar, whereas blue ones are based on the assumption of timing noise only in pulsars where it was reported in~\cite{goncharov2020pptadr2noise}.
    Right: Common red-noise parameter estimation with the free-spectral model. Lines represent the full PPTA data, whereas filled regions represent PPTA DR2 without PSR J0437$-$4715.  The black line is the inferred spectrum assuming a power-law model with $\gamma=13/3$.  Vertical dotted lines represent inverse orbital periods of solar system planets.}
    \label{fig:crn2}
\end{figure*}

Following the process described above, we obtained strong evidence for ${\rm CP1}$, with $\ln \mathcal{B}^{\text{CP1}}_{\varnothing} > 15.0$.
This implies an odds ratio in favor of CP1 of $> 3\times 10^6:1$, if both models had even odds {\em a priori}.   
The results from the parameter estimation are shown in the  left panel in Figure~\ref{fig:crn2}. We obtain\footnote{Unless otherwise specified, throughout the paper uncertainties provide 1-$\sigma$ credible levels.}  $\log_{10}A_{\text{CP1}} = -14.55^{+0.10}_{-0.23}$ and $\gamma = 4.11^{+0.52}_{-0.41}$.  The results from the NANOGrav analysis are overlaid and have significant overlap, although the NANOGrav analysis preferred a steeper spectral exponent, when the latter analysis was conducted with $5$ Fourier components.  Unlike in \cite{arzoumanian2020gwb}, our measurements are consistent when we changed the numbers of fluctuation frequencies used to model the common process.

The right panel in Figure~\ref{fig:crn2} represents the parameter estimation for the free-spectral model.  It is challenging to obtain a complete noise model for the brightest MSP,  PSR~J0437$-$4715~\citep{goncharov2020pptadr2noise}, and so we show this spectrum with, and without, the inclusion of this pulsar.  
We overlay the astrophysically interesting spectrum corresponding to $\gamma = 13/3$, along with the frequencies corresponding to the orbital periods of the planets.

\begin{figure*}[!htb]
\includegraphics[width=9cm]{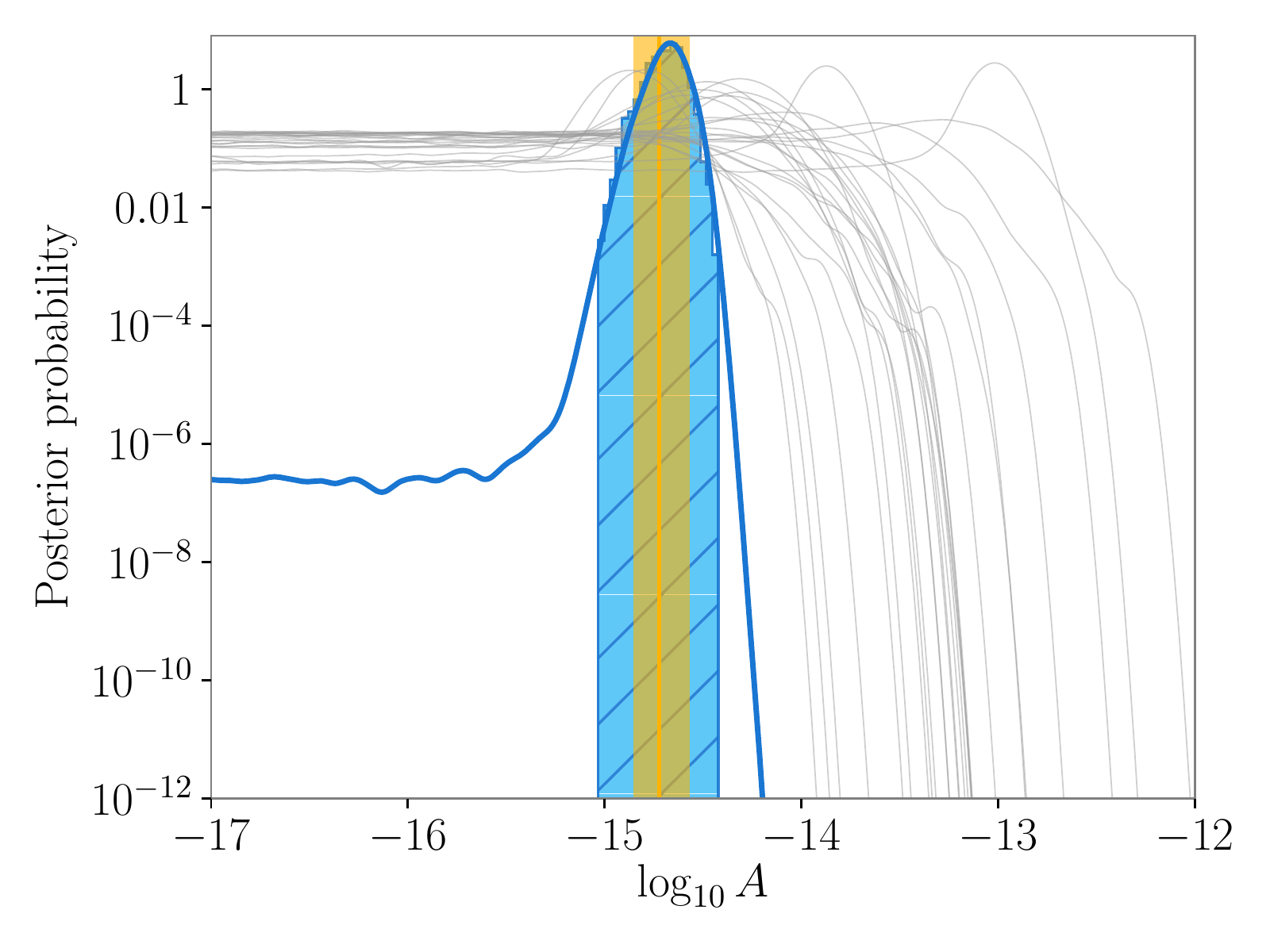}
\includegraphics[width=9cm]{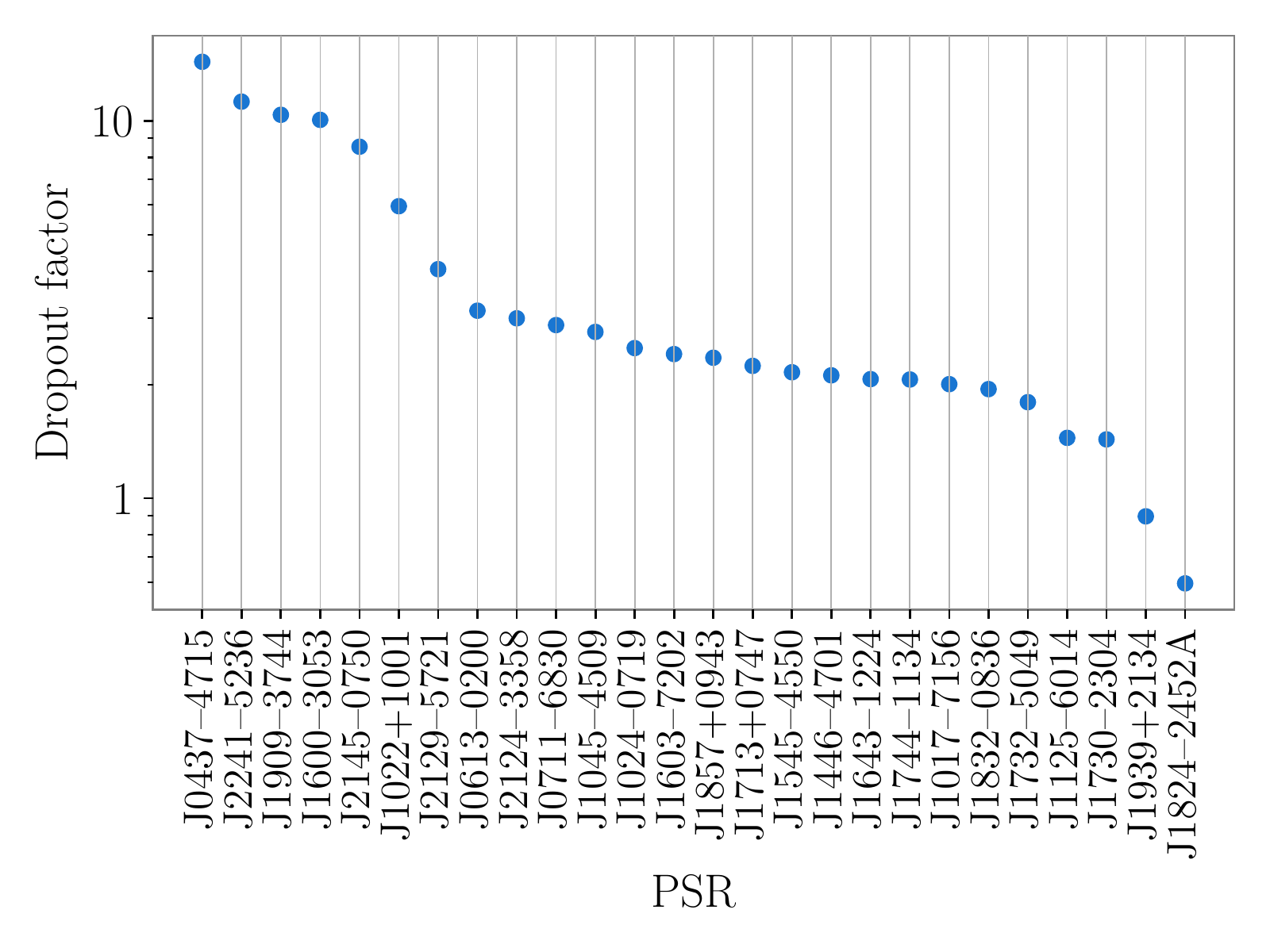}
    \caption{Pulsar contributions to the common red noise, assuming a fixed power-law index of $-13/3$ (${\rm CP2}$). Left: posterior distributions for the common red-noise amplitude, $A$. The hatched blue area is the result of a joint analysis of all pulsars with fixed white-noise parameters. The thick blue line shows the distribution obtained from a  factorized likelihood approach. Thin grey lines show contributions from individual pulsars to the factorized posterior.  The yellow vertical line and the shaded region represent the median and 1-$\sigma$ levels of the NANOGrav measurement. Right: Dropout factors for PPTA DR2 pulsars. We interpret the dropout factors to represent the consistency of noise in a given pulsar with ${\rm CP2}$, as discussed in Section \ref{sec:dropout}.}     
    \label{fig:crn1}
\end{figure*}

In the left-hand panel of Figure~\ref{fig:crn1}, we show the posterior distribution for $\log_{10}A$, assuming a power-law model with $\gamma = 13/3$. The measured ${\rm CP2}$ amplitude is $\log_{10}A = -14.66 \pm 0.07$ corresponding to $A = 2.2^{+0.4}_{-0.3} \times 10^{-15}$, which is consistent with that measured in the NANOGrav 12.5-year data set. 

The NANOGrav analysis attempted to determine which pulsars contributed to this signal by calculating  dropout factors, which for our sample are displayed in the right-hand panel of  Figure~\ref{fig:crn1}.  The pulsars with the smallest dropout factors (PSRs J1824$-$2452A and J1939$+$2134) are  known to have strong timing noise inconsistent in strength with that of other pulsars.  However the pulsars with the highest dropout factors include pulsars with high (PSR~J0437$-$4715) and low timing precision (PSR J1022$+$1001), and pulsars with shorter timing baselines (PSR~J2241$-$5236).     The meaning and use of dropout factors is further discussed in Section~\ref{sec:dropout}.

\begin{figure*}
    \begin{tabular}{cc}
    \includegraphics[width=0.48\textwidth]{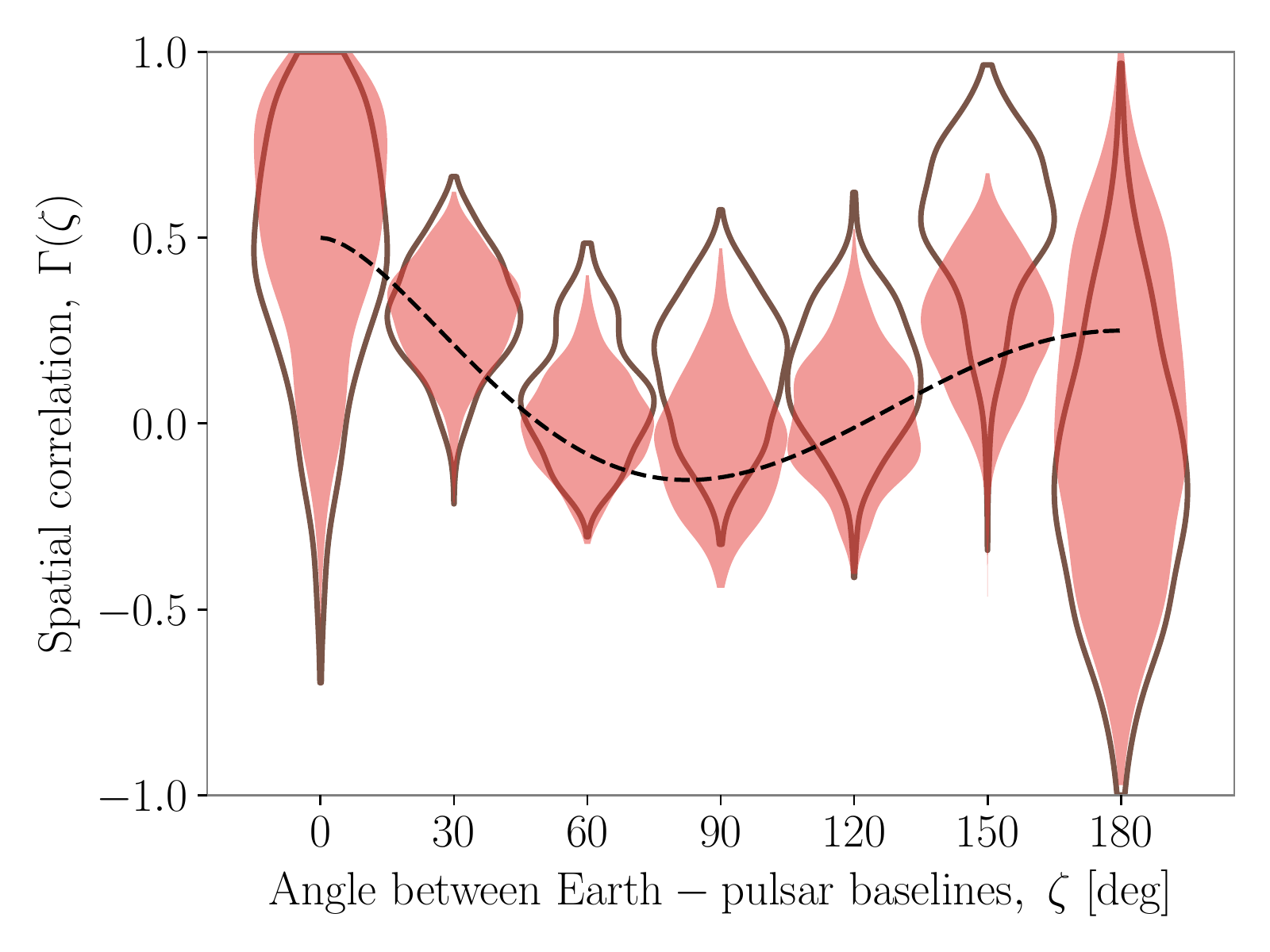}
 &     \includegraphics[width=0.48\textwidth]{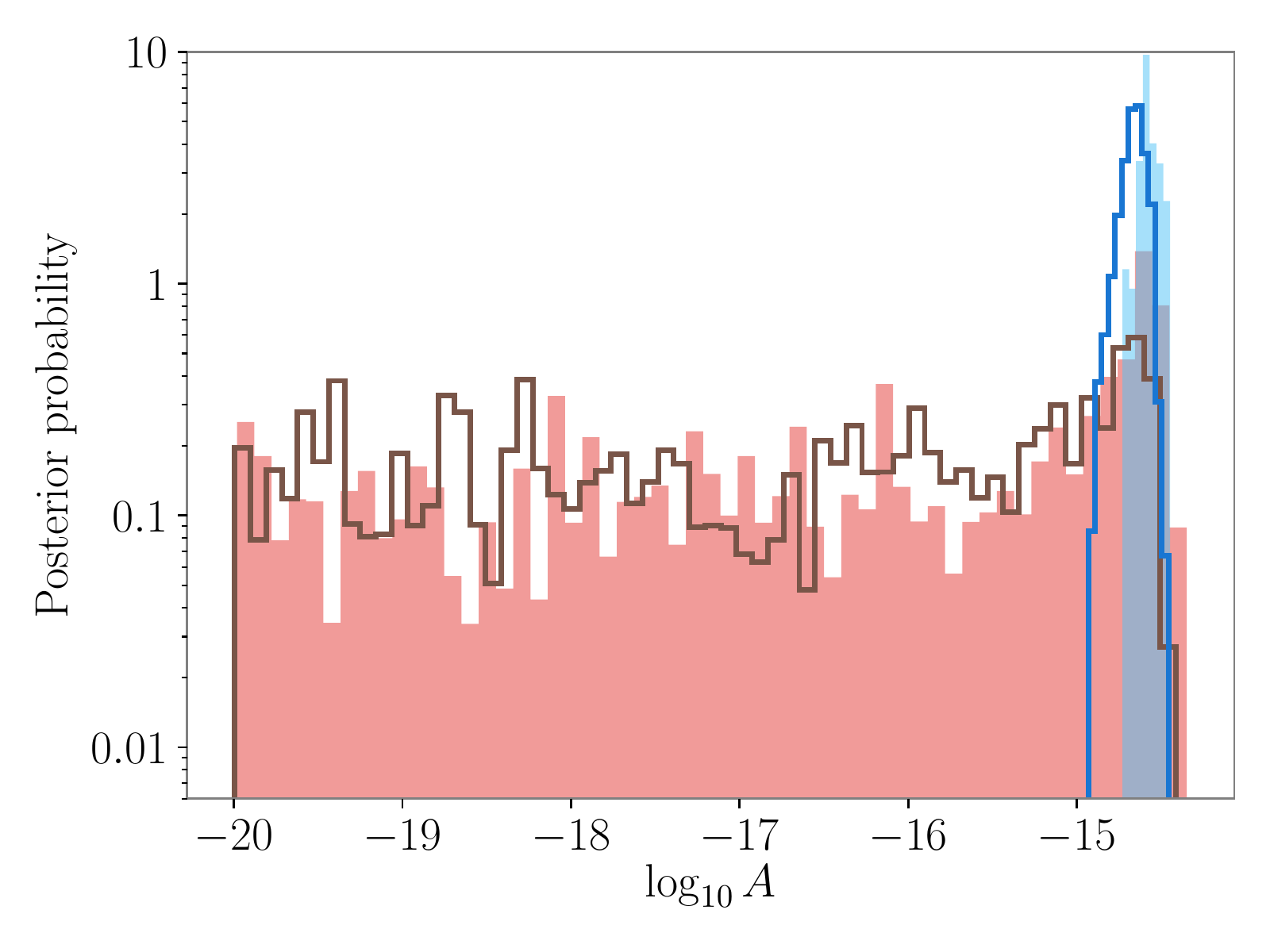} \\
\end{tabular}
    
    \caption{
    Left:  Inferred inter-pulsar spatial correlations in PPTA DR2 at seven node angles $\zeta$ between Earth-pulsar baselines.
    The dashed line is the predicted correlation from the gravitational-wave background. 
    Right: Power-law amplitude of the Hellings-Downs process without auto-correlation (red) and of the common red noise (blue). 
    In both figures, lines represent the full PTA based on the assumptions of timing noise only in pulsars according to~\cite{goncharov2020pptadr2noise}, whereas the filled regions represent PPTA DR2 without PSR~J0437$-$4715 and timing noise terms in all pulsars.
    }
    \label{fig:orfhd}
\end{figure*}

The results from the model-independent parameter estimation of the overlap reduction function (obtained assuming $\gamma = 13/3$) are provided in Figure~\ref{fig:orfhd}.  
The left-hand panel shows the inferred spatial correlations.
They were sampled at seven node angles, whereas spatial correlations for other angles were obtained with linear interpolation.
The Hellings-Downs relation is overplotted.
The right-hand panel shows inferred amplitude of the  Hellings-Downs spatially correlated noise and that for ${\rm CP2}$.   
With the entire PPTA sample of pulsars the Bayes factor is $\ln \mathcal{B}^{\text{HD}}_{\text{CP2}}=0.3$, which provides no significant evidence for, or against, Hellings-and-Downs correlations. We note that if PSR~J0437$-$4715 is removed from the sample then the Bayes factor increases to $\ln \mathcal{B}^{\text{HD}}_{\text{CP2}}=1.0$.   The data strongly disfavor ${\rm CP2}$ having monopole or dipole spatial correlations ($\ln \mathcal{B}^\text{MP}_\text{CP2}$ and $\ln \mathcal{B}^\text{DP}_\text{CP2}$ both $< -10$).

\section{Discussion}\label{sec:discussion}

Under the same assumptions the Bayesian analyses of both the PPTA and NANOGrav data sets show a preference for models which include a common noise process in addition to individual noise terms.  The ${\rm CP1}$ model has consistent spectral index and amplitude between the data sets and therefore we can exclude this signal from being telescope dependent.  Given the different strategies employed in mitigating  interstellar propagation effects by the two projects (both in terms of choice of observing band and methods for correcting for dispersion-measure variations), it is also unlikely that the noise is associated with the interstellar medium.

However, we are  attempting to detect a common noise process from a single realization of the process in each pulsar.   The noise process is strongest at lowest fluctuation frequency, so the process is being characterised on a same time scale comparable to the typical data span.
This greatly complicates tests of the noise modelling.  
Consequently there are a number of caveats for interpreting the ${\rm CP1}$ and ${\rm CP2}$ results as discussed in the following sub-sections.  We conclude by discussing our search for spatial correlations in the data.

\subsection{Are the models of the intrinsic noise correct and complete?}

We have modeled the intrinsic noise to be a power-law process.  
Intrinsic timing noise for millisecond pulsars is not well studied over the relevant time scales.  We know millisecond pulsars  \cite[PSRs~J0613$-$0200 and J1824$-$2452A][]{cognard_1824glitch,mckee_0163glitch} have exhibited glitch events, so  small glitches are possibly present in other pulsars.
Two of the pulsars in our sample, PSRs~J0437$-$4715 and J2241$-$5236, have reported evidence for excess non-stationary noise \cite[][]{goncharov2020pptadr2noise}.
Large-scale studies of non-millisecond pulsars have demonstrated that the amplitude of their timing noise is approximately determined by the pulsar spin-down rate~\citep{aditimingnoise}, but there is also clear evidence for discrete changes in spin frequency or frequency derivative \cite[][]{cordes_timingnoise}, which may occur at quasi-periodic intervals~\cite[][]{hobbsquasi}.  It is therefore unlikely that the intrinsic pulsar timing noise is perfectly modeled via a power-law process.  We know that the pulse profiles of millisecond pulsars can show secular shape changes \cite[][]{shannon_magnetosphere}, which, if unmitigated, result in a non-stationary noise process. 


    

\subsection{What assumptions lead to the evidence of a common process?}    
    
The assessment of whether a common red-noise process is present is based on the null hypothesis that all pulsars are affected by independent red timing noise, modeled by a power law spectrum described by amplitudes and exponents drawn from a uniform prior.
The detection hypothesis is that all pulsars also exhibit a red process with the same power-law spectrum, where amplitudes and exponents are drawn from the delta-function prior.
There is a possibility that neither of these models nor their priors are sufficient descriptions of the data, which is  often referred to as model misspecification \cite[e.g.][]{specification,misspec_muller}.
In simulations we demonstrate that noise without a statistically identical spectrum between pulsars can be misinterpreted as the common red process.

\begin{figure*}
    \centering
    \includegraphics[width=0.49\textwidth]{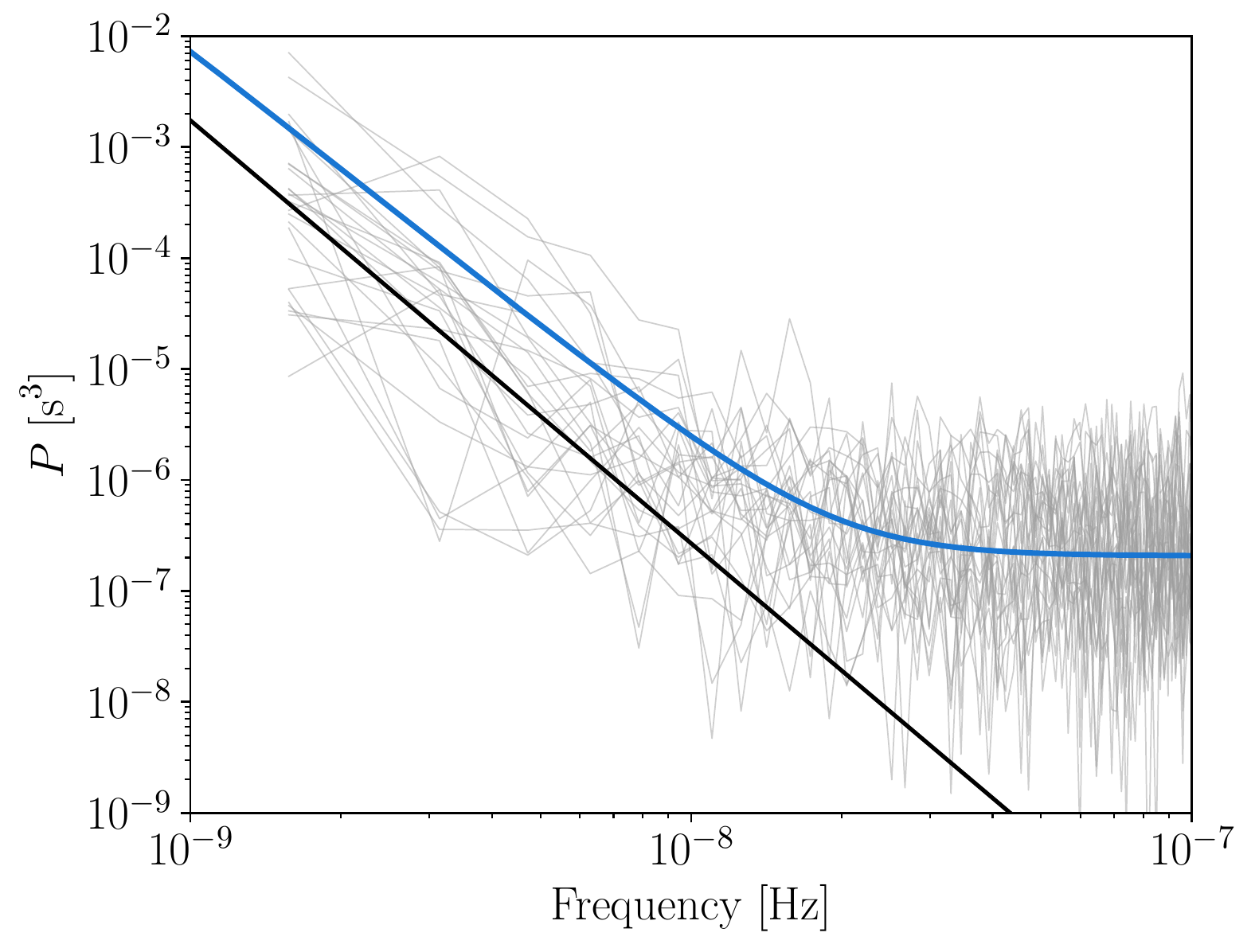}
    \includegraphics[width=0.49\textwidth]{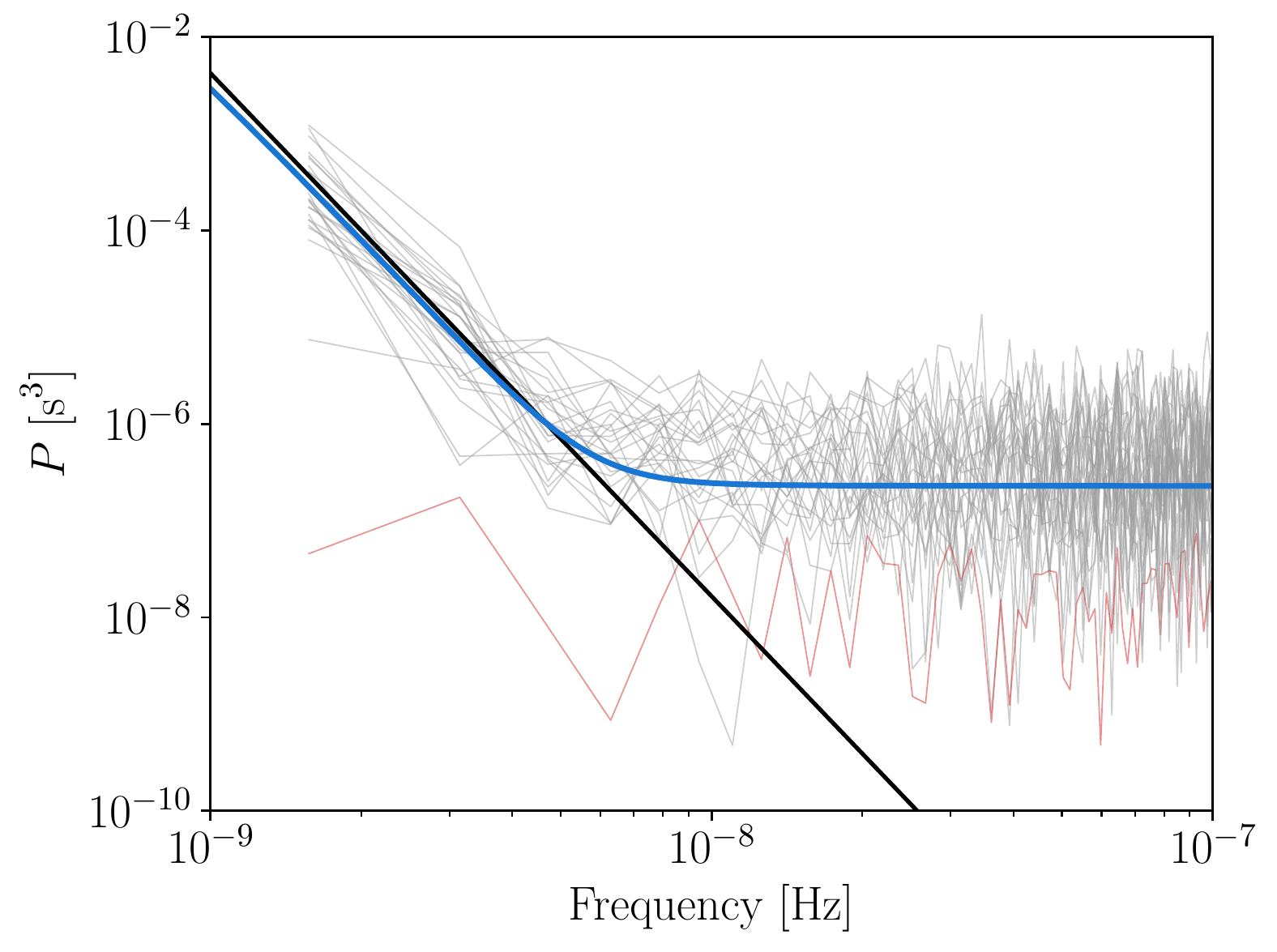}
    \caption{\label{fig:sim_psd}Power spectral density of the two sets of simulations that mimic a common red process. The grey lines represent the injected input noise spectra of simulated timing residuals for the 26 pulsars. The spectra were formed using a generalized least squares technique  \cite[][]{coles_cholesky}. The blue lines represent the pulsar-averaged values, and the black lines indicate the recovered common spectrum. Left: Pulsars with similar, but not identical, timing noise properties.
    Right: A simulation where 25 pulsars have identical timing noise properties, and one has a lower and shallower timing noise than the others (red line).
    }
\end{figure*}

\begin{figure}
    \centering
    \includegraphics[width=8.5cm]{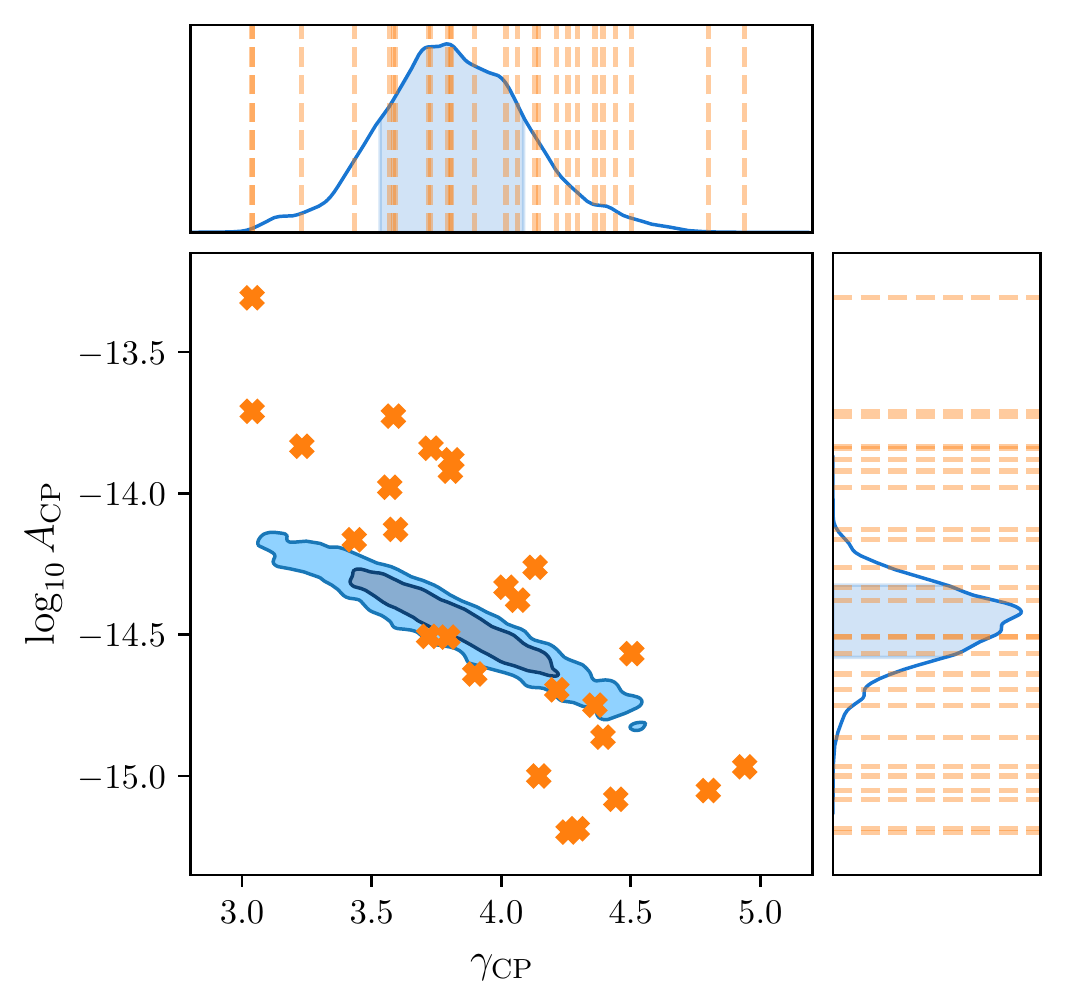}
    \caption{\label{fig:similar_A_gamma}Simulated steep timing noise parameters and recovered common red process parameters, corresponding to the left panel in Figure~\ref{fig:sim_psd}. The contours represent posterior probability density for the common-process amplitude and spectral index. The amplitudes and spectral indices of the injected timing-noise spectra for the 26 pulsars are indicated by the orange crosses and orange dashed lines. }
\end{figure}

We simulated 10 realizations of timing residuals for the 26 PPTA-DR2 pulsars, with a range of realistic white noise levels, and injected power-law timing noise models with amplitudes and spectral indices drawn uniformly across approximately several orders of magnitude ($\log_{10} A$ spanning approximately $-16$ to $-13$ and $\gamma$ spanning $3$ to $5$).
The power spectral densities of the simulated residuals are shown in the left-hand panel of Figure~\ref{fig:sim_psd}. 
We performed model selection for a model with a common-spectrum process along with intrinsic pulsar timing noise ($\rm{CP2}$), against a model with intrinsic timing noise only ($\varnothing$), and obtained $\ln\mathcal{B}^{\rm{CP2}}_{\varnothing} > 13.5$ in all realizations, implying that our methodology can detect a ``common'' process if the properties of the noise are broadly similar but far from identical, with the amplitude of the noise varies by three orders of magnitude.  
Figure~\ref{fig:similar_A_gamma} shows the recovered common noise spectrum and the injected timing noise models.
We continued to increase the spread in injected noise amplitudes upward from $\approx 3$ orders of magnitude, and found that common noise is disfavoured when the spread in amplitude exceeds 5 orders of magnitude.
Thus, the simulated data only favors the correct hypothesis when the range of simulated noise parameters starts to resemble the uniform red noise prior assumed in recent analyses.

It is physically likely that intrinsic pulsar timing noise has similar, but not identical properties between different pulsars \cite[][]{shannon_timingnoise}.  As such, a second null hypothesis (not tested by the current analysis) would be that each pulsar has independent red noise, but the properties of that noise cluster in a similar range.  
This could be examined by assuming that the amplitude and spectral index of the noise terms for the pulsars are not identical, but are drawn from a distribution.
The non-uniform noise hypothesis is distinct both from the signal hypothesis with the delta-function prior and from the noise hypothesis with the uniform prior.
For example, the new noise prior could be modeled as a Gaussian distribution of width $\mu_A$ and $\mu_\gamma$, and variance $\sigma_A^2$ and $\sigma_\gamma^2$.
If the variance is inconsistent with zero it would suggest that the noise processes were not common but just similar. 
If it is consistent with zero then values could be used to constrain the properties of a common process.

As a gravitational-wave background signal will affect all pulsars, it will be apparent not only as a common spectral process, but also as a ``noise floor''.  The noise level of a given pulsar should not be below this floor apart from statistical fluctuations, including instances where two noise processes cancel each other out.   We updated the simulations shown in Figure~\ref{fig:sim_psd} by including timing noise with identical power-law spectral densities  ($\log_{10} A = -15.51$, $\gamma = 5.5$) for 25 pulsars, and a lower, shallower-spectrum timing noise ($\log_{10} A = -14.38$, $\gamma = 1.5$) in the pulsar with the lowest white noise levels.
The power spectral densities corresponding to these simulations are shown in the right-hand panel of Figure~\ref{fig:sim_psd}. We performed model selection for $\rm{CP2}$ over $\varnothing$, and again found significant support for the common-spectrum noise process ($\ln\mathcal{B}^{\rm{CP2}}_{\varnothing} >$ $13.1$ across all realizations), which is comparable to the support found in both the NANOGrav and PPTA analyses. As the power spectral density in the lowest frequency channel for the low-noise pulsar is $4$ orders of magnitude lower than the common signal, the model selection is not explicitly identifying a noise floor.
However, a factorized-likelihood dropout analysis showed that the simulated pulsar with a low noise level was not consistent with the retrieved common-spectrum process (with a dropout factor $<1$; see the following section). 

\subsection{Do we understand which pulsars are contributing to the evidence?}\label{sec:dropout}

The  dropout factor does not explicitly assess individual pulsar contributions to a common-noise process. As shown in Figure~\ref{fig:crn1}, unexpected pulsars (such as PSR~J2241$-$5236, which is a relatively new addition to the PPTA sample) have high dropout factors. As the dropout factor is the integral of the product of an individual posterior distribution constraint on the common red process with the posterior distribution of the apparent common-spectrum process in the rest, pulsars with uninformative posterior distributions (i.e., pulsars with relatively high white noise and no evidence of red noise) can still have significant dropout factors. Therefore, the dropout factor represents not a ``contribution statistic'', but a ``consistency statistic'':  pulsars with high factors have noise that is not inconsistent with the presence of a common process.  If we wish to determine which pulsars contribute to the evidence for a red-noise process then we could calculate evidence values in favor of common noise, starting from the pulsar which provides the best (or worst) single pulsar limit on the gravitational-wave background, and  iteratively increasing the number of pulsars, or calculate the change in evidence when removing individual pulsars from the array.  

\subsection{Are we affected by the choice of solar system ephemeris?}

The amplitude of ${\rm CP2}$ is higher than the two 95\% confidence upper limits previously set by NANOGrav ($A < 1.45 \times 10^{-15}$) and the PPTA  ($A < 1.0 \times 10^{-15}$).
The PPTA limit was based on the DE421 Solar System ephemeris, without marginalizing over errors in the ephemeris and included data up to the beginning of 2015.

In order to carry out an initial exploration, to see if the results in this paper may be affected by errors within the solar system ephemeris, we modeled a subset of the potential errors as parametrized perturbations in the ephemerides~\citep[\textsc{bayesephem,}][]{vallisneri2020sse}. We performed model selection for perturbations in (1) the masses of Mars, Jupiter and Saturn (2) the individual Keplerian orbital elements of these planets and (3) in the rate of rotation about the ecliptic pole. These terms, except for Mars and Saturn, were marginalized over by~\cite{arzoumanian2020gwb}.
The resulting Bayes factors are negative and hence we conclude that the presence of such errors is not favored by the data.
We also performed the same analysis using the JPL DE438, DE430 and DE421 ephemerides.  Neither DE436 nor DE438 provide evidence for any errors.
Using DE430, we find positive $\ln \mathcal{B} = 1.6$ for an error in one of the Jupiter's orbital elements and $\ln \mathcal{B} = 0.6$ favoring both the error in Saturn's mass and one of the Saturn's orbital elements.  Using the DE421 ephemeris, the oldest one that we tested, we only find a positive $\ln \mathcal{B} = 0.3$ for an error in the mass of Saturn.


%

\subsection{Searching for spatial correlations}

Only a correlation analysis will provide incontrovertible evidence of a gravitational-wave background and we currently have no statistical evidence for the presence of spatial correlations. 
The overlap reduction function in Figure~\ref{fig:orfhd} is consistent with the expected correlations from a gravitational-wave background (in particular if PSR~J0437$-$4715 is removed from the analysis).
We note that the bins in Figure~\ref{fig:orfhd} are interpolated and correlated, which may boost the apparent significance seen by eye. 
As shown in Figure~\ref{fig:crn1}, the noise spectrum in PSR~J0437$-$4715 is consistent with ${\rm CP2}$ and hence it provides the highest dropout factor.
However, as shown in the right-hand panel of Figure~\ref{fig:orfhd}, the inclusion of PSR~J0437-4715 lowers the posterior probability of spatial correlations at the maximum-posterior value of $A$ of ${\rm CP2}$, further diminishing the evidence for a gravitational-wave background.
Unfortunately, the only telescope with a long timing baseline for PSR~J0437$-$4715  is  Parkes and it is therefore  challenging to confirm the noise modelling for this pulsar.
In the future it will be possible to compare with observations taken with, for example, MeerKAT as part of the MeerTime project \citep{bailes2020}, or with the Jansky Very Large Array as part of NANOGrav.

When evidence for Hellings-Downs correlations in pulsar timing array data sets is found, it will be important to examine the hypothesis tested.  Simulations, such as ones containing uncorrelated red noise could be used to determine the likelihood of improperly modeled uncommon red noise inducing these spatial correlations.

With the provisos given above we have no evidence that the detected ${\rm CP1}$ or ${\rm CP2}$ noise process is linked to a gravitational-wave background.  However, if the signal is a bona-fide astrophysical gravitational-wave background, then the relatively high amplitude would favor high merger-rate densities, short merger timescales, and high normalisations for the black hole - galaxy bulge mass relation \citep{middleton2020ngresults}.
The near-future prospects for probing the origin of the signal and the underlying dynamics of the supermassive black hole population are discussed in~\cite{pol2020milestones}.
\cite{sesana2013gwbamp} showed that background amplitudes of $> 10^{-15}$ could be caused by the effect of overmassive black holes on black hole - host relations.
The detected amplitude is also within observationally-constrained limits based on the local supermassive black-hole mass function~\citep{zhu2019minmaxgwb}.

\section{Conclusions}


Under the assumptions of an analysis of the NANOGrav 12.5-year data set~\citep{arzoumanian2020gwb}, we have detected with high confidence a common-spectrum time-correlated signal in the timing of the 26 PPTA-DR2 millisecond pulsars.
However, as noted above, there are some important caveats that need to be addressed before the signature could be confidently attributed to a physical process common to all the pulsars in the array.
We do not confirm or rule out that the common-spectrum process is a spatially-correlated stochastic gravitational-wave signal.
However, the identified process does not possess monopole or dipole correlations and is not caused by errors in masses and trajectories of Mars, Jupiter and Saturn, that would have resulted in deterministic timing residuals according to \textsc{bayesephem} models~\citep{vallisneri2020sse}.
Proposed follow-up work relates to (1) improving our understanding of the properties of the intrinsic timing noise in millisecond pulsars and (2) identifying the optimal model comparisons and methodologies that can determine whether the noise detected corresponds to a ``noise floor'' and is identical in all pulsars.

The PPTA project now has nearly three additional years of data  obtained with a higher sensitivity wide-band system \cite[][]{hobbsuwl} that can be added  with the data presented to increase timing baselines. 
We will also combine the PPTA data sets with observations from other observatories as part of the International Pulsar Timing Array project \cite[][]{perera_iptadr2}, with the latter being ideal to continue this work. Such lengthened and more sensitive data sets will allow us to probe time scales significantly longer than the orbital period of Jupiter and closer to that of Saturn. We will be able to compare noise models obtained from a wide range of telescopes and maximise our chance of determining the nature of the red-noise signals that are present in our data.


These data sets will also enable more sensitive and robust searches for the Hellings-Downs spatial correlations necessary to make a definitive detection of the gravitational-wave background.





\section*{Acknowledgements} \label{sec:acknowledgements}
This work has been carried out by the Parkes Pulsar Timing Array, which is part of the International Pulsar Timing Array.
The Parkes radio telescope (Murriyang) is part of the Australia Telescope, which is funded by the Commonwealth Government for operation as a National Facility managed by CSIRO. This paper includes archived data obtained through the CSIRO Data Access Portal (\href{http://data.csiro.au}{data.csiro.au}).
We acknowledge the use of \textsc{chainconsumer}~\citep{chainconsumer}. 
Parts of this research were conducted by the Australian Research Council  (ARC) Centre of Excellence for Gravitational Wave Discovery (OzGrav), through project number CE170100004.  RMS acknowledges support through ARC future fellowship FT190100155.  RS acknowledges support through the ARC Laureate fellowship FL150100148.
The author list is based on three tiers, which correspond to primary contributors, to members of the collaboration who provided feedback, and to members of the collaboration with significant observing record.
The data-processing code that was used in this work is available at \href{https://github.com/bvgoncharov/correlated_noise_pta_2020/}{github.com/bvgoncharov/correlated\_noise\_pta\_2020}.

\bibliography{mybib}{}
\bibliographystyle{aasjournal}

\end{document}